# Theory of the nanoparticle-induced frequency shifts of whispering-gallery-mode resonances in spheroidal optical resonators

L. Deych* and V. Shuvayev

*Physics Department, Queens College of CUNY, Flushing, NY 11367, USA*

**Corresponding author:* lev.deych@qc.cuny.edu

Nanoparticle-induced modifications of the spectrum of whispering-gallery-modes (WGM) of optical spheroidal resonators are studied theoretically. Combining an *ab initio* solution of a single resonator problem with a dipole approximation for the particle, we derive simple analytical expressions for frequencies and widths of the particle-modified resonances, which are valid for resonators with moderate deviations from the spherical shape. The derived expressions are used to analyze spectral properties of the resonator-particle system as functions of the particle's position, the size of the resonators and the characteristics of WGMs. The obtained results are shown to agree well with available experimental data. It is also demonstrated that the particle-induced spectral effects can be significantly enhanced by careful selection of resonator's size, refractive index and other experimental parameters. The results presented in the paper can be useful for applications of WGM resonators in biosensing, cavity QED, optomechanics and others.

## I. INTRODUCTION

Modes of optical whispering-gallery-mode (WGM) resonators are characterized by large quality (Q) factors and small mode volume, which makes them an attractive candidate for a variety of applications [1,2]. One of the important properties of WGMs is their sensitivity to changes in the surrounding medium - even a small subwavelength object placed in the proximity of the resonator's surface can significantly affect their spectrum [3–7] and spatial profile of the modes [5,6,8]. The modification of the spectrum of WGMs due to interaction with nanoparticles has attracted a great deal of attention in connection with the quest for optical methods of single particle detection and analysis [9–23], which recently resulted in a label-free single protein detection [12,13], and was used for tracking individual atoms in cavity QED experiments [24].

The typical spectral modification due to a nanoparticle consists in "splitting" of a single WGM resonance into a doublet of resonances. While doublets of "split" WGM resonances have been observed a while ago and were ascribed to backscattering due to surface rougness of the resonator [25–28], demonstration of this effect due to a single nano-sized scatterer was achieved only in Ref. [4], where a tip of an optical near-field microscop served as a nano-sized perturbation. Ability to control the position of the scatterer allowed the authors of Ref. [4] to demonstrate that, contrary to previous views, the "splitting" observed in this and earlier experiments was not symmetric, with two scattering-induced resonances appearing on both sides of the initial resonance. It was shown instead that the observed doublet consisted of an initial resonance, not affected by the particle, and a new red-shifted particle-induced resonance. Later these results were reproduced and used for detection and sizing of real particles in a number of subsequent papers [14–19]. Sometimes, in the case of resonators with lower Q-factors interacting with smaller particles, the split resonances cannot be resolved and the overlapping peaks appear as a single resonance red-shifted from its position in the absence of the particle [9–13,21]. This regime is particulalry important for biosensing applications whose main goal is to achieve label-free single protein detection [3,10–13,29,30].

The theoretical description of the spectral charactersitics of the particle-WGM resonator systems lags far behind the experimental achievements. It relies primarily on euristic "Reactive Sensing Principle" (RSP) [3] believed to describe well the regime of overlapping resonances, the phenomenological model of Ref. [4], which treat a WGM resonator as a system of two degenerate modes coupled by the particle to a common bath of propogating free space modes, and *ab initio* calculations of Ref. [5–7,23] dealing with perfectly spherical [5–7] and two-dimensional disk resonators [23].

Since in this paper we are mostly interested in the regime of resolved resonances, we shall leave a discussion of the RSP for the next publication while focusing here on the clockwise-counterclockwise mode coupling model of Ref. [4] and *ab initio* calculations of Ref. [5–7]. The main difference between the model presented in Ref. [4] and earlier attempts to describe splitting of WGMs [27] is the idea that the particle does not couple degenerate WGMs directly, but instead couples them to a common bath of propagating modes. This feature of the model allowed authors of Ref. [4] to explain experimentally observed perseverance of the initial WGM resonance even in the presence of the particle and to derive a simple expression for the particle-induced splitting of the initial resonance. At the same time, the phenomenological nature of this model, based on the number of rather arbitrary assumptions, does not allow for verification of the limits of its applicability and leaves open the question about completeness of its description of the phenomenon under consideration.

The *ab initio* results presented in Ref. [5–7], while providing a complete and rigorous description of the problem, are directly applicable only to the idealized situation of the spherical resonators, and it is not clear *a priori* to what extent the obtained results can be used for real resonators, which always deviate from spherical shape. Comparison of the predictions of Ref. [5–7] with experimental data shows that the theoretical results significantly exceed the experimentally observed frequency splitting, indicating that the spherical approximation does not describe real nominally spherical resonators very well [31].

A significant progress in understanding the WGM-nanoparticle interaction could be achieved by extending the *ab initio* approach to non-spherical, in particular, spheroidal resonators. Resonators of this shape are of particular interest since their properties are expected to be closest to those of spherical resonators, but mainly because the spheroidal shape is a good approximation for nominally spherical resonators used in many experimental studies. This problem, however, is rather difficult and has not yet been addressed either analytically or numerically. The reason for this is the lack of appropriate symmetries, which would allow reducing the dimensionality of the problem and simplify both computational and analytical calculations. The only known to us previous attempt to study such a system involved numerical modeling of a toroidal resonator interacting with a nanoparticle based on the Finite Element Method [32]. To make this problem numerically tractable the authors of that work invoked an additional assumption that the particle affects the field of the resonator only in its immediate vicinity. While such an approach can give a perturbative correction to the field of the resonator due to the particle, it is not able to predict modification of the frequencies of WGM resonances, which is the main interest for many applications.

Recently we developed an approach allowing to get around the above-mentioned difficulties and to devise a theoretical framework yielding rigorous quantitative description of optical resonances in the spheroidal resonator interacting with a subwavelength polarizable object. This approach is based on combining the T-matrix formalism used to describe field of a single resonator [33] with the dipole approximation for the field of the nanoparticle. We shall present the detailed description of this approach and the results of rigorous numerical simulations of the resonator-particle system based on it elsewhere. In this paper we use this framework in conjunction with a number of well-controlled approximations to derive simple analytical results for the particle-induced frequencies and widths of the respective resonances. The analysis of the derived expressions discovers the features of the resonator-particle interaction missed by the phenomenological approaches and establishes the limits of their applicability. The developed theory also explains how a deviation from the spherical shape affects the spectral properties of the resonators and shows a very good agreement with available experimental data.

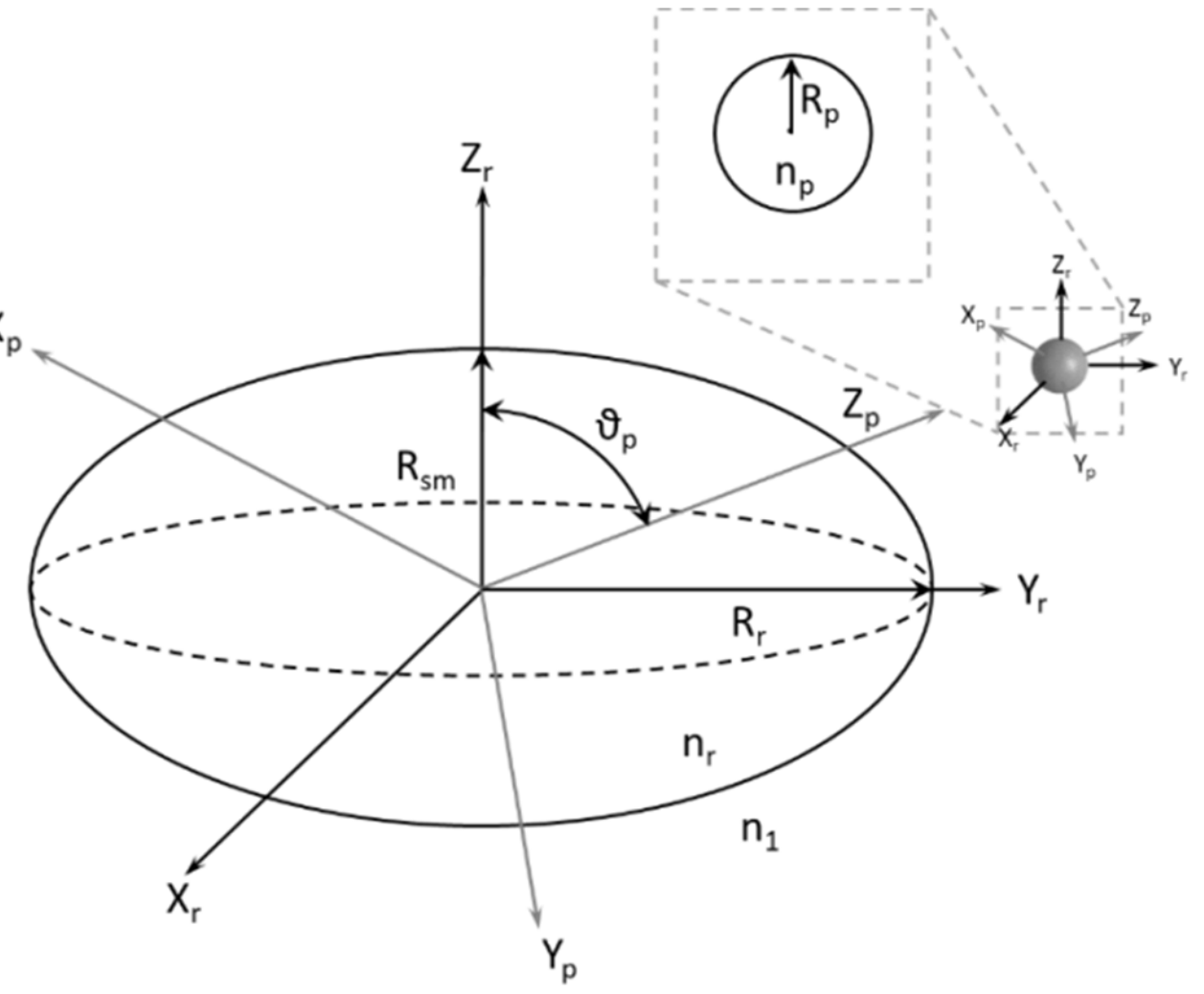

FIG. 1. Resonator-particle system with all geometric and material parameters. Also shown are resonator's and particle's coordinate systems centered at the resonator and the particle. The magnified image of the particle in the right corner shows its size and refractive index.

## II. RESONANT APPROXIMATION: DERIVATION

We consider a spheroidal resonator, characterized by an equatorial radius $R_r$, small radius $R_{sm}$, refractive index $n_r$, interacting with a subwavelength particle of radius $R_p$ and refractive index $n_p$ positioned in its vicinity (see Figure 1). It is assumed that the entire system is surrounded by a dielectric medium with refractive index $n_1$. We describe the interaction between the resonator and the particle by combining the T-matrix formalism, used to describe the field of a single resonator [33,34], with the dipole approximation for the field of the nanoparticle. The T-matrix approach is based on presenting the fields incident on the resonator and scattered by it as linear combinations of Vector Spherical Harmonics (VSH) [33] of transverse electric (TE), $\mathbf{M}_{m,l}^{(1,3)}(k_1 r_r, \theta_r, \varphi_r)$, and transverse magnetic (TM), $\mathbf{N}_{m,l}^{(1,3)}(r_r, \theta_r, \varphi_r)$, polarizations:

$$\begin{aligned} \mathbf{E}_{inc}^{(r)} &= \sum_{l,|m|\le l} \left[ a_{m,l} \mathbf{M}_{m,l}^{(1)}(k_1 r_r, \theta_r, \varphi_r) + b_{m,l} \mathbf{N}_{m,l}^{(1)}(k_1 r_r, \theta_r, \varphi_r) \right] \\ \mathbf{E}_{sc}^{(r)} &= \sum_{l,|m|\le l} \left[ c_{m,l} \mathbf{M}_{m,l}^{(3)}(k_1 r_r, \theta_r, \varphi_r) + g_{m,l} \mathbf{N}_{m,l}^{(3)}(k_1 r_r, \theta_r, \varphi_r) \right] \end{aligned}, \quad (1)$$

where $r_r, \theta_r, \varphi_r$ are radial, polar and azimuthal coordinates defined in a particular spherical coordinate system, and subscript $r$ points out that the origin of this system is at the resonator's center. Electric field described by the VSH of TE polarization is characterized by vanishing radial component, and in the case of VSH of TM polarization it is the respective magnetic field, which lacks its radial component. Superscripts (1) and (3) in $\mathbf{M}_{m,l}^{(1,3)}, \mathbf{N}_{m,l}^{(1,3)}$ indicate that the radial dependence of these VSHs is given by spherical Bessel functions $j_l(k_1 r)$ or outgoing spherical Hankel functions $h_l^{(1)}(k_1 r)$ of the first kind, respectively. Parameter $k_1$ is defined as $k_1 = n_1 k$, where $k$ is the vacuum wave number of electromagnetic field with frequency $\omega$. While the field scattered by the resonator includes combination of VSHs of both polarizations with all possible values of modal indexes $m$ and $l$: $l > 1, -l \le m \le l$, the field scattered by the subwavelength particle is described as a linear combination of only TM VSHs with $l = 1$:

$$\mathbf{E}_{sc}^{(p)} = \sum_{|m|\le 1} p_m \mathbf{N}_{m,1}^{(3)}\left(k_1 r_p, \theta_p, \varphi_p\right), \quad (2)$$

where subscript $p$ indicates that the origin of the respective spherical coordinate system is now at the center of the particle. Eq. (2) constitutes the dipole approximation for the particle's field.

The modal indexes $l, m$ in Eq. (1) and (2) characterize spherical harmonics defined with respect to a particular spherical coordinate system. While in atomic physics the modal index $m$ is usually called the magnetic number, in the optical context we will call it, following Ref. [35], a polar number, while index $l$ will be referred to as an orbital number. Our formalism is based on utilization of the coordinate systems with different directions of their polar axis $Z$. One, which we shall call the resonator's system, has its polar axis $Z_r$ directed along the axis of symmetry of the resonator, while the other, which will be referred to as the particle's system, is characterized by the polar axis $Z_p$ connecting the centers of the resonator and the particle (see Figure 1). Representation of the same field in terms of VSHs defined in different coordinate system is obviously different. Transformation between these two systems can be performed via rotation by Euler angles

$$\alpha = \pi / 2; \beta = -\vartheta_p; \gamma = 0 \ , \tag{3}$$

where we adopt notations for the Euler angles from Ref. [33], and $\vartheta_p$ is the polar coordinate of the center of the particle in the resonator's coordinate system. Both coordinate systems can be centered either at the center of the resonator (resonator-centered) or the center of the particle (particle-centered). Transition between the resonator-centered and particle-centered systems with parallel polar axis is given by vectors $\pm\mathbf{d}_{pr}$, where $\mathbf{d}_{pr}$ is a position vector of the particle with respect to the center of the resonator.

In addition to angular modal numbers $l, m$, WGMs are also characterized by a radial index, $s$, determining the radial behavior of the field of a WGM at the resonance frequency.

All the information about the field scattered by the system, including the positions and the widths of the resonances, is contained in the coefficients $c_{m,l}, g_{m,l}$ and $p_m$. In the T-matrix formalism the expansion coefficients of the scattered field $c_{m,l}, g_{m,l}$ are related to the respective coefficients of the incident field by a T-matrix $T^{(\sigma,\sigma')}_{m,l;\mu,\nu}$ [33], where upper indexes correspond to two different polarizations of the VSH with $\sigma = 1$ chosen to represent TE polarization, while $\sigma = 2$ represents the TM polarized VSHs. In the resonator's coordinate system the T-matrix of a spheroidal resonator is diagonal with respect to polar indexes $m, \mu$: $\tilde{T}^{(\sigma,\sigma')}_{m,l;\mu,\nu} = \tilde{T}^{(\sigma,\sigma')}_{m,l;m,\nu}\delta_{m,\mu}$ [33], where we used tilde to indicate that the T-matrix is defined in the resonator's coordinates. The remaining off-diagonal components of the T-matrix describe cross-polarization and cross-modal scattering. For weakly non-spherical objects these components are relatively small and can be neglected, while the diagonal components $\tilde{T}^{(\sigma,\sigma)}_{m,l;m,l}$ in a vicinity of an isolated WGM resonance can be approximated as

$$\tilde{T}^{(\sigma,\sigma)}_{l,m;l,m} \approx -\frac{i\gamma^{(\sigma)}_{l,m,s}}{\omega-\omega^{(\sigma)}_{l,m,s}+i\gamma^{(\sigma)}_{l,m,s}} , \tag{4}$$

where $\omega^{(\sigma)}_{l,m,s}$ and $\gamma^{(\sigma)}_{l,m,s}$ are frequency and linewidth of the respective resonance characterized by orbital, polar and radial indexes $l,\ m,\ \mathrm{s}$, respectively, and polarization $\sigma$. Eq. (4) has the same form as the single resonance approximation for the Mie-Lorentz coefficients describing WGM resonances in ideal spheres. The only difference between the two is dependence of $\omega^{(\sigma)}_{l,m,s}$ and $\gamma^{(\sigma)}_{l,m,s}$ in Eq. (4) upon the polar number $m$. This dependence reflects lifting of the $2l+1$ -fold degeneracy of WGMs when complete spherical symmetry of spherical resonators is lowered to an axial symmetry of spheroids. The presentation of the diagonal elements of the T-matrix in the form of Eq. (4) is justified by numerical computations of these elements using codes publicly available at [36]. One can see from Figure 2 that in both spherical and spheroidal cases the real part of the diagonal elements is equal to -1 at the resonance, justifying presentation of these elements in the form of Eq. (4). At the same time, a single resonance in a spherical resonator is replaced by several resonances differing by their polar numbers $m$ in the spheroid and they are now significantly shifted with respect to each other. One can notice that the peak corresponding to $m=l$ (fundamental mode) only slightly deviates from the respective peak in the spherical resonator. In the case of the radial mode of the first order, the position of the fundamental resonance in the spheroidal resonator would have been even closer to the resonance of the spherical resonator.

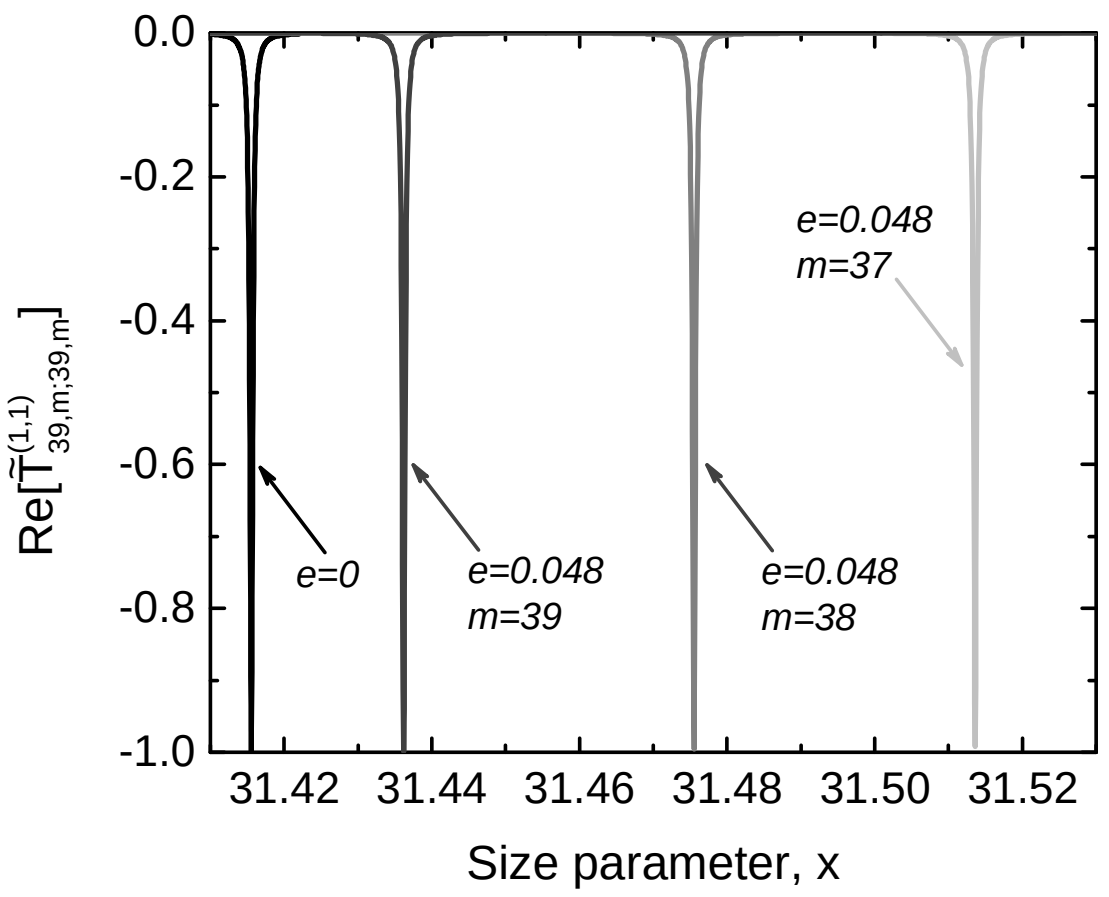

FIG. 2. Frequency dependence of the element of T-matrix in the vicinity of the second order radial resonance ($s$=2) expressed in terms of the size parameter $x = kR_r$. The black curve shows this dependence for a spherical resonator ($e$=0), for which peaks with the same $l$, but different $m$, coincide. The gray curves show the same graphs for spheroidal resonator with ellipticity parameter $e = 1 - R_{\mathrm{sm}} / R_r = 0.048$ for different values of $m$.

The scattering coefficients $c_{m,l}$, $g_{m,l}$ and $p_m$ depend on the incident field used to excite the resonances. We consider excitations of two types assuming that the incident field is such that in *a spherical resonator* it would have excited a field described in the **resonator's coordinate system** by a single VSH of either TE or TM polarization with a given modal numbers $l = L$, $m = M$, and radial number $s = S$. We distinguish between the types of excitation by introducing superscript $\sigma$ in the notations for the coefficients of the incident and scattered fields in Eq. (1) and (2): $a_{m,l}^{(\sigma)}, c_{m,l}^{(\sigma)}, g_{m,l}^{(\sigma)}, p_m^{(\sigma)}$. The value of $\sigma = 1$ corresponds again to the TE type of excitation, and $\sigma = 2$ corresponds to the excitation of the TM type. In the **resonator's coordinate system** the excitation of TE and TM types is introduced by choosing the coefficients of the incident field in the form

$$\begin{aligned} \tilde{a}_{m,l}^{(1)} &= a_0 \delta_{m,M} \delta_{l,L}; \tilde{b}_{m,l}^{(1)} = 0 \\ \tilde{a}_{m,l}^{(2)} &= 0; \tilde{b}_{m,l}^{(2)} = b_0 \delta_{m,M} \delta_{l,L} \end{aligned} \tag{5}$$

and assuming that the frequency $\omega$ of the exciting field is in the vicinity of the respective WGM resonance $\omega_{L,M,S}^{(\sigma)}$. Parameters $a_0, b_0$ in Eq. (5) characterize the intensity of the exciting radiation and can be determined from the experimental value of the power entering the resonator. These assumptions approximate the realistic excitation of WGMs with the field of a tapered fiber [37,38], in which one neglects excitations of non-resonant parasitic modes and the influence of the fiber on the frequencies and Q-factors of the excited resonances.

In the presence of a particle interacting with the resonator we have to deal with the two-body problem, which we treat by generalizing the multi-sphere Mie theory [39,40] to the case of non-spherical scatterers. In the spirit of this approach the field scattered by the resonator plays the role of the incident field for the particle, which is assumed to have a spherical shape. The field scattered by the particle, in its turn, combines with the external illuminating field and, therefore, contributes to the expansion coefficients of the incident field. Using the T-matrix approach to relate the expansion coefficients in Eq. (1) to those of the modified incident field as well as the vector addition theorem [41] for transforming VSHs defined in the resonator-centered to the particle-centered coordinate system and vice versa, we derive the system of equations for coefficients $p_m^{(\sigma)}, m = -1, 0, 1$ in Eq. (2). The vector addition theorem describes transformation of VSHs upon translation of the origin of the coordinate system used for their definition. VSHs defined in the resonator-centered coordinate system can be presented in terms of VSHs defined in the particle-centered system as:

$$\begin{aligned}\mathbf{N}_{m,l}^{(3)}\left(k_1r_r,\theta_r,\varphi_r\right)=\sum_{\nu=1}^{\infty}\sum_{\mu=-\nu}^{\nu}\Big[&A_{\mu,\nu;m,l}^{(+)}\left(k_1,-\mathbf{d}_{pr}\right)\mathbf{N}_{\mu,\nu}^{(1)}\left(k_1r_p,\theta_p,\varphi_p\right)+\\&B_{\mu,\nu;m,l}^{(+)}\left(k_1,-\mathbf{d}_{pr}\right)\mathbf{M}_{\mu,\nu}^{(1)}\left(k_1r_p,\theta_p,\varphi_p\right)\Big]\\\mathbf{M}_{m,l}^{(3)}\left(k_1r_r,\theta_r,\varphi_r\right)=\sum_{\nu=1}^{\infty}\sum_{\mu=-\nu}^{\nu}\Big[&A_{\mu,\nu;m,l}^{(+)}\left(k_1,-\mathbf{d}_{pr}\right)\mathbf{M}_{\mu,\nu}^{(1)}\left(k_1r_p,\theta_p,\varphi_p\right)+\\&B_{\mu,\nu;m,l}^{(+)}\left(k_1,-\mathbf{d}_{pr}\right)\mathbf{N}_{\mu,\nu}^{(1)}\left(k_1r_p,\theta_p,\varphi_p\right)\Big]\end{aligned}, \qquad (6)$$

where $A_{m,\nu;\mu,l}^{(+)}(k_1,\mathbf{d}_{pr})$ and $B_{m,\nu;\mu,l}^{(+)}(k_1,\mathbf{d}_{pr})$ are the translation coefficients, which can be found, for instance, in Ref. [33]. These coefficients have important symmetry properties with respect to inversion of the translation vector $\mathbf{d}_{pr}$ and interchange of the modal indexes [42]

$$\begin{aligned}A_{m,1;\mu,\nu}^{(+)}(k_1,-\mathbf{d}_{pr})&=\left[A_{\mu,\nu;m,1}^{(-)}(k_1,\mathbf{d}_{pr})\right]^*\\B_{m,1;\mu,\nu}^{(+)}(k_1,\mathbf{d}_{pr})&=\left[B_{\mu,\nu;m,1}^{(-)}(k_1,\mathbf{d}_{pr})\right]^*\end{aligned}. \qquad (7)$$

The superscripts $(+)$ or $(-)$ in the translation coefficients indicates that their radial dependence is given by the spherical Hankel function of the 1st or the 2nd kind respectively. In the **particle's coordinate system** the translation coefficients are diagonal in polar indexes $m,\mu$ making this system the most convenient for the purpose of this work. Thus, the rest of the calculations are made using the particle's coordinate system, in which equations for coefficients $p_m^{(\sigma)}$ can be presented as:

$$\sum_{\mu=-1}^{1}\left[\delta_{m,\mu}-\alpha_p\left(U_{m,\mu}+V_{m,\mu}\right)\right]p_\mu^{(\sigma)}=\alpha_p\left(k_1R_p\right)\sum_l\left[c_{l,m}^{(\sigma,0)}B_{m,1;m,l}^{(+)}\left(k_1,-\mathbf{d}_{pr}\right)+g_{l,m}^{(\sigma,0)}A_{m,1;m,l}^{(+)}\left(k_1,-\mathbf{d}_{pr}\right)\right]. \qquad (8)$$

Parameter $\alpha_p\left(k_1R_p\right)$ in Eq. (8) is the Lorenz - Mie coefficient [33] describing relations between the dipole $(l=1)$ components of the incident and the scattered TM fields for a spherical particle of radius $R_p$

$$\alpha_p(x) = -\frac{n_1^2 j_1(n_1 x)\left[n_p x\, j_1(n_p x)\right]' - n_p^2 j_1(n_p x)\left[n_1 x\, j_1(n_1 x)\right]'}{n_1^2 h_1(n_1 x)\left[n_p x\, j_1(n_p x)\right]' - n_p^2 j_1(n_p x)\left[n_1 x\, h_1(n_1 x)\right]'}\Bigg|_{x=kR_p}, \tag{9}$$

where $\left[zf(z)\right]'$ means differentiation over the entire argument of the respective function, $c_{l,m}^{(\sigma,0)}$ and $g_{l,m}^{(\sigma,0)}$ are the expansion coefficients of the resonator's field in Eq. (1) in the absence of the particle. Also, thanks to the dipole approximation, only translation coefficients with one of the orbital index $l=1$ appear in Eq. (8). In this case general formulas for the translation coefficients significantly simplify, yielding

$$\begin{gathered} A_{\mu,\nu;\mu,1}^{(+)}(k_1,\mathbf{d}_{pr}) = \sqrt{\frac{3}{2}}\left[\sqrt{\frac{(\nu+1)(\nu+|\mu|)}{(2\nu+1)(1+|\mu|)}}h_{\nu-1}^{(1)}(k_1 d_{pr}) + (-1)^{\mu}\sqrt{\frac{\nu(\nu+1-|\mu|)}{(2\nu+1)(1+|\mu|)}}h_{\nu+1}^{(1)}(k_1 d_{pr})\right]. \\ B_{\mu,\nu;\mu,1}^{(+)}(k_1,\mathbf{d}_{pr}) = \mathrm{i}\frac{\sqrt{3}}{2}\mu\sqrt{2\nu+1}h_{\nu}^{(1)}(k_1 d_{pr}) \end{gathered} \tag{10}$$

In the small vicinity of a WGM with frequency $\omega_{L,m,S}^{(\sigma)}$, the main contribution to the properties of the system comes from the elements of the T-matrix $T_{m,L;\mu,L}^{(\sigma,\sigma)}$, which can be found by using transformation properties of the T-matrix with respect to rotations. For rotations described by Euler angles given by Eq. (3), the transformation between resonator's and particle's coordinate systems is given by [33]

$$T_{m,L;\mu,L}^{(\sigma,\sigma)} = \sum_{m_1=-L}^{m_1=L} (-1)^{\mu-m_1} d_{m,m_1}^{(L)}\left(\vartheta_p\right)\tilde{T}_{m_1,L;m_1,L}^{(\sigma,\sigma)} d_{m_1,\mu}^{(L)}\left(\vartheta_p\right), \tag{11}$$

where $d_{m,M}^{(L)}(\vartheta_p)$ is the d-(small) Wigner matrix, an explicit expression for which can be found, for example, in Ref. [33]. In most cases of practical interest the particle induced frequency shift is much smaller than the spectral distance between frequencies $\omega_{L,m,S}^{(\sigma)}$ with different values of polar number $m$, split by the deviation of the resonator's shape from spherical. In such a case one can leave in Eq. (11) only two terms with $m_1 = \pm M$, which correspond to two WGMs of a single resonator often called clockwise and counterclockwise propagating WGMs [4], which remain degenerate even in spheroidal resonators. The polar number $M$ here is fixed by the frequency of the exciting field. In this approximation the T-matrix in the particle's coordinates can be written down as

$$T^{(\sigma,\sigma)}_{m,L;\mu,L} = \left[1+(-1)^{\mu+m}\right] d^{(L)}_{m,M}\left(\theta_p\right) d^{(L)}_{\mu,M}\left(\theta_p\right) \tilde{T}^{(\sigma,\sigma)}_{M,L;M,L} . \qquad (12)$$

If one neglects all elements of the T-matrix except of those given in Eq. (11) and (12), the matrices $U_{m,\mu}$ and $V_{m,\mu}$ in Eq. (8) take the form:

$$\begin{aligned} U_{m,\mu} &= T^{(2,2)}_{m,L;\mu,L} A^{(+)}_{\mu,L;\mu,1}(k_1, \mathbf{d}_{pr}) A^{(+)}_{m,1;m,L}\left(k_1, -\mathbf{d}_{pr}\right) \\ V_{m,\mu} &= T^{(1,1)}_{m,L;\mu,L} B^{(+)}_{\mu,L;\mu,1}(k_1, \mathbf{d}_{pr}) B^{(+)}_{m,1;m,L}\left(k_1, -\mathbf{d}_{pr}\right) \end{aligned} . \qquad (13)$$

Finally, parameters $c^{(\sigma,0)}_{l,m}$ and $g^{(\sigma,0)}_{l,m}$ in Eq. (8) can be found in the particle's coordinate system by applying transformation properties of the VSHs [33] to Eq. (5). Neglecting, again, non-diagonal in polarization indexes elements of the T-matrix, they can be presented as

$$\begin{aligned} c^{(1,0)}_{l,m} &= (-i)^M \, d^{(l)}_{m,M}(\vartheta_p) \tilde{T}^{(1,1)}_{M,l;M,L} a_0; \quad c^{(2,0)}_{l,m} = 0 \\ g^{(2,0)}_{l,m} &= (-i)^M \, d^{(l)}_{m,M}(\vartheta_p) \tilde{T}^{(2,2)}_{M,l;M,L} b_0; \quad g^{(1,0)}_{l,m} = 0 \end{aligned} . \qquad (14)$$

Once particle's coefficients $p^{(\sigma)}_m$ are found from Eq.(8), the non-vanishing expansion coefficients $c^{(1)}_{m,l}, g^{(2)}_{m,l}$ of the resonator's field can be easily determined from

$$\begin{aligned} c^{(1)}_{m,l} &= c^{(1,0)}_{l,m} + \sum_{\mu=-1}^{1} T^{(1,1)}_{m,l;\mu,l} B^{(+)}_{\mu,l;\mu,1}(k_1, \mathbf{d}_{pr}) p^{(1)}_\mu \\ g^{(2)}_{m,l} &= g^{(2,0)}_{l,m} + \sum_{\mu=-1}^{1} T^{(2,2)}_{m,l;\mu,l} A^{(+)}_{\mu,l;\mu,1}(k_1, \mathbf{d}_{pr}) p^{(2)}_\mu \end{aligned} . \qquad (15)$$

Eq. (8) through (15) constitute the resonant approximation for the problem under consideration. It is similar to the resonant approximation introduced in Ref. [5,6] for spherical resonators, and is valid if the ellipticity of the resonator is small enough to validate assumptions expressed by Eq. (13) and large enough to validate Eq. (12). Numerical simulations and experimental data indicate that these conditions are fulfilled for most situations of practical interest. In the rest of this paper we discuss properties of the resonator-particle system as described by this approximation and compare them with the results of previous theories [4] and available experimental data [19].

## III. RESONANT APPROXIMATION: RESULTS

### A. TE polarization

Since in the resonant approximation we neglect the cross-polarization scattering, resonances of the resonator-particle system can be characterized as being of TE or TM types excited by the external field of the respective polarization. In this subsection we present the results for the TE case, described by Eq. (8) with index $\sigma$ set to 1. Taking into account Eq. (13), we see that in the vicinity of a TE resonance we can neglect contribution to Eq. (8) from matrix $U_{m,\mu}$, so that in the resonant approximation particle's field coefficients $p_m^{(1)}$ are determined by matrix $V_{m,\mu}$, whose elements with $m=0$ or $\mu=0$ vanish (see Eq. (10)). As a result, only coefficients $p_{\pm1}^{(1)}$ are different from zero in this case, and one can derive for the resonator's field coefficients for an arbitrary value of particle's angular coordinate $\theta_p$:

$$c_{m\pm,L}^{(1)}\left(\theta_p\right)=\begin{cases}\dfrac{(-i)^M a_0\left[d_{m,M}^{(L)}\left(\theta_p\right)\pm d_{-m,M}^{(L)}\left(\theta_p\right)\right]}{\left[\tilde{T}_{M,L;M,L}^{(1)}\right]^{-1}+\alpha_p\left[d_{1,M}^{(L)}\left(\theta_p\right)-d_{-1,M}^{(L)}\left(\theta_p\right)\right]^2\left[B_{1,L;1,1}^{(+)}(k_1,d_{pr})\right]^2}\\[2ex] \dfrac{(-i)^M a_0\left[d_{m,M}^{(L)}\left(\theta_p\right)\pm d_{-m,M}^{(L)}\left(\theta_p\right)\right]}{\left[\tilde{T}_{M,L;M,L}^{(1)}\right]^{-1}+\alpha_p\left[d_{1,M}^{(L)}\left(\theta_p\right)+d_{-1,M}^{(L)}\left(\theta_p\right)\right]^2\left[B_{1,L;1,1}^{(+)}(k_1,d_{pr})\right]^2}\end{cases}, \tag{16}$$

where

$$c_{m\pm,L}^{(1)}=c_{m,L}^{(1)}\pm c_{-m,L}^{(1)} . \tag{17}$$

For coefficients with even values of $m$, $c_{m+,L}^{(1)}$ is given by the first line of Eq. (16) with "+" sign chosen, and $c_{m-,L}^{(1)}$ is given by the second line with "-" sign chosen. This designation is reversed for the coefficients with odd values of the polar number: $c_{m-,L}^{(1)}$ is now given by the first line of Eq. (16) with "-" sign chosen, while $c_{m+,L}^{(1)}$ is found from the second line with "+" sign chosen. According to Eq.(16), the response of the resonator-particle system is characterized by two resonance frequencies defined by the poles of the respective expressions. Using Eq.(4), the frequencies and the respective widths of the resonances can be presented as

$$\begin{aligned} \delta\omega_{L,M,S}^{(1,\pm)} &= -\left(d_{1,M}^{(L)} \pm d_{-1,M}^{(L)}\right)^2 \gamma_{L,M,S}^{(1)} \operatorname{Im}\left(\alpha_p \left[B_{1,L;1,1}^{(+)}(n_1 k_{L,M,S}, d_{pr})\right]^2\right) \\ \delta\gamma_{L,M,S}^{(1,\pm)} &= -\left(d_{1,M}^{(L)} \pm d_{-1,M}^{(L)}\right)^2 \gamma_{L,M,S}^{(1)} \operatorname{Re}\left(\alpha_p \left[B_{1,L;1,1}^{(+)}(n_1 k_{L,M,S}, d_{pr})\right]^2\right) \end{aligned}, \tag{18}$$

where $\delta\omega_{L,M,S}^{(1,\pm)}$ describe shifts of particle-induced resonant frequencies from the initial single-resonator resonance, $\delta\gamma_{L,M,S}^{(1,\pm)}$ is the particle-induced broadening of the resonance, and $k_{L,M,S}$ is computed at the respective unperturbed frequency $\omega_{L,M,S}$. Thus, in the generic case, both members of the particle-induced doublet are shifted from the initial resonance, contrary to the result of Ref. [4]. Therefore, it is no longer appropriate to describe this situation in terms of "frequency splitting" defined as a spectral distance between the two members of the doublet. Instead, we will discuss it in terms of the shifts of each individual particle-induced resonance.

For equatorial position of the particle, $\theta_p = \pi/2$, Wigner's d-matrices acquire additional symmetry $d_{-m,M}^{(L)}(\pi/2) = (-1)^{L-M} d_{m,M}^{(L)}(\pi/2)$. As a result, one of the frequency shifts ($\delta\omega_{L,M,S}^{(1,-)}$ for even $L-M$ and $\delta\omega_{L,M,S}^{(1,+)}$ for odd $L-M$) and the respective width modifications vanish, meaning that the position and the width of one of the resonances in the presence of the particle coincides with those of the initial WGM. Also, for the particle at the equatorial position, either $c_{m+,L}^{(1)}$ or $c_{m-,L}^{(1)}$ (depending upon the sign of $(-1)^{L-M}$) turns zero, and one obtains for the $c_{m,L}^{(1)}$:

$$c_{m,L}^{(1)} = \begin{cases} 2(-i)^M a_0 d_{m,M}^{(L)}(\pi/2) \tilde{T}_{M,L;M,L}^{(1)} \\ \dfrac{2(-i)^M a_0 d_{m,M}^{(L)}(\pi/2)}{\left[\tilde{T}_{M,L;M,L}^{(1)}\right]^{-1} + 4\alpha_p \left[d_{1,M}^{(L)}(\pi/2)\right]^2 \left[B_{1,L;1,1}^{(+)}(k_1, d_{pr})\right]^2} \end{cases}, \tag{19}$$

where polar number $m$ takes even values in the first line and odd values in the second line for even values of $L-M$, and the parity of $m$ in different lines of Eq. (19) is reversed for odd $L-M$. Apparently, the coefficients in the first line of Eq. (19) resonate on the frequency of the initial WGM. It also follows from this result that for $\theta_p = \pi/2$ the resonator's field coefficients with opposite signs of the polar number $m$ are related according to $c^{(1)}_{m,L} = (-1)^{L-M} c^{(1)}_{-m,L}$. Combining this with the known properties of the VSH: $\mathbf{X}_{l,m}(\theta,\varphi) = (-1)^m \left[\mathbf{X}_{l,-m}(\theta,\varphi)\right]^*$, where symbol $\mathbf{X}$ stands for the angular part of a VSH of any polarization, we can conclude that the modes corresponding to the resonances described by Eq. (19) have definite parity with respect to azimuthal angle $\varphi$ (defined in the particle's coordinate system): for even values of $L-M$ the field of the non-shifted resonance is even in the angle $\varphi$, while the field of the particle-induced resonance is odd in it; for odd values of $L-M$ the situation is reversed. This separation of the modes of the resonator-particle system into odd and even functions of $\varphi$ is a consequence of the symmetry of the system with respect to reflection in the equatorial plane of the resonator, which is preserved when the particle is positioned in this plane.

This symmetry also helps to understand why the initial single-resonator frequency survives in the presence of the particle for the TE polarized modes. Indeed, consider electric field at the location of the particle. In the case of TE polarization, this field, which does not have a radial component, is normal to the polar axis of the particle's coordinate system. Operation of reflection in the plane $X_r Y_r$ acts on the field in two ways: it changes the sign of its non-zero components as well as the sign of the azimuthal angle in the VSHs representing the field. As a result, the field composed of the even with respect to $\varphi$ functions changes its sign, and must, therefore, vanish at the location of the particle due to symmetry of the system with respect to this reflection. Accordingly, the resonance associated with these components of the field does not "notice" the presence of the particle. On the other hand, the field composed of the odd functions remains invariant with respect to the reflection. Respectively, the symmetry does not require it to vanish and these components resonate on a new particle-induced frequency.

Comparison of Eq. (18) with respective results of Ref. [5–7] for spherical resonators shows that in the resonant approximation the main effect of the deviation of the resonator from sphere is factor $\left(d^{(L)}_{-1,M} \pm d^{(L)}_{1,M}\right)^2$, which is mainly responsible for dependence of the resonant frequencies on the particle's angular coordinate $\theta_p$ (additional dependence on $\theta_p$ comes from the variation of the resonator-particle distance $d_{pr}$ with the angle in spheroidal resonators, but this dependence is very weak for the ellipticity parameters of interest and can be neglected). In the case of spherical resonators this factor is absent and splitting depends, therefore, only upon radial coordinate of the particle $d_{pr}$. The Wigner d-function $d_{1,M}^{(L)}$ decreases very fast with increasing orbital number $L$ and is, therefore, responsible for reduction of the splitting in spheroidal resonators as compared to the spherical case.

In order to illustrate the obtained results we compute the dependence of frequency shifts $\delta\omega_{L,M,S}^{(1,\pm)}$ on particle's angular coordinate $\theta_p$ and orbital number $L$ for several special situations. For the purpose of these calculations we assumed that the positions $\omega_{L,M,S}^{(1)}$ and widths $\gamma_{L,M,S}^{(1)}$ of the single resonator WGMs appearing in Eq. (18) can be found neglecting the deviation of the resonator's shape from spherical. This assumption was verified by comparison with numerical calculations using T-matrix and was found to work well for modes with polar numbers such that $L-M \ll L$. This allows us to compute $\omega_{L,M,S}^{(1)}$ and $\gamma_{L,M,S}^{(1)}$ from the poles of the spherical Lorentz-Mie coefficients rather than from diagonal elements of the T-matrix. This assumption is not of principle nature and was used only to reduce the amount of numerical efforts. Should more accurate computations be necessary, we can always determine these parameters directly from the T-matrix.

We begin by considering a resonator-particle system with parameters close to those, which were studied experimentally in Ref. [19]: $R_r = 43.5\,\mu m$, $n_r = 1.44$; $R_p = 50\,nm$, and $n_p = 1.59$. We also take into account that in Ref. [19] the resonator and the particles were placed in aqueous environment so that we choose the refractive index of the surrounding medium to be $n_1 = 1.33$. The frequency of the laser used in the experiments of Ref. [19] belonged to $600\,nm$ band, which roughly corresponds to a TE WGM with orbital number $L = 642$ and radial number $S = 1$, or to orbital number $L = 630$ and radial number $S = 2$. We focus our attention to the particle-induced frequency shifts, while using the expression for the modified linewidth to verify that the shifted resonances remain well separated in all situations under consideration.

In Figure 3 we present the results of the calculations of the frequency shifts assuming equatorial position of the particle for WGMs with orbital number extending from $L=160$ to $L=650$, radial numbers $S=1$, $S=2$ and polar numbers $M=L,\ M=L-2$ (the mode with $M=L-1$ has zero splitting for the equatorial position of the particle). The most remarkable feature of this plot is a crossover between the first and second radial modes: for smaller values of the orbital number the first radial mode demonstrates the larger splitting, while for larger values of $L$ the splitting in the second radial mode prevails.

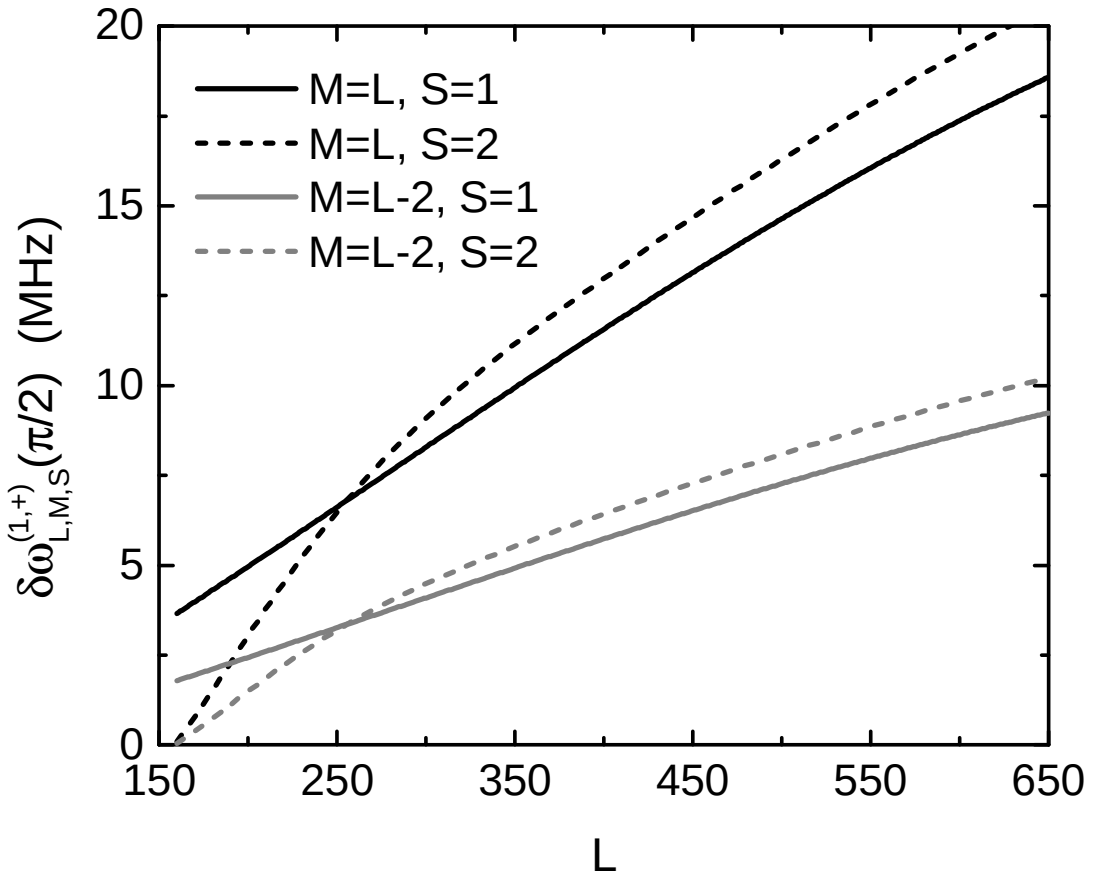

FIG. 3. Dependence of the frequency shifts $\delta\omega_{L,M,S}^{(1,+)}(\pi/2)$ versus orbital number $L$ for the equatorial position of the particle. The two upper graphs correspond to $M=L$ mode and radial numbers $S=1$ and $S=2$, while the two lower curves represent $M=L-2$ mode with the same two radial numbers. The second shift $\delta\omega_{L,M,S}^{(1,-)}(\pi/2)$ vanishes for the equatorial position of the particle.

This feature is common to both the fundamental and $M=L-2$ modes, while the splitting of the latter is significantly smaller. Such behavior is the result of competition of two opposite tendencies. On one hand, the magnitude of the field at the position of the particle is smaller for the second radial mode compared to the first one, but on the other hand, the radiative rate $\gamma_{L,M,S}^{(1)}$, which appears in Eq. (18), is larger for the second mode whose Q-factor is smaller. Apparently, for large enough $L$, the second of these tendencies prevails over the first one, resulting in the observed crossover.

In Figure 4 we present the dependence of $\delta\omega_{L,L,S}^{(1,+)}$ and $\delta\omega_{L,L-1,S}^{(1,-)}$ for $L=630$, $M=L,\ L-1$ (single and double peak lines, respectively) and $S=1,2$ upon particle's angular coordinate $\theta_p$. As expected, the largest frequency shift in the fundamental

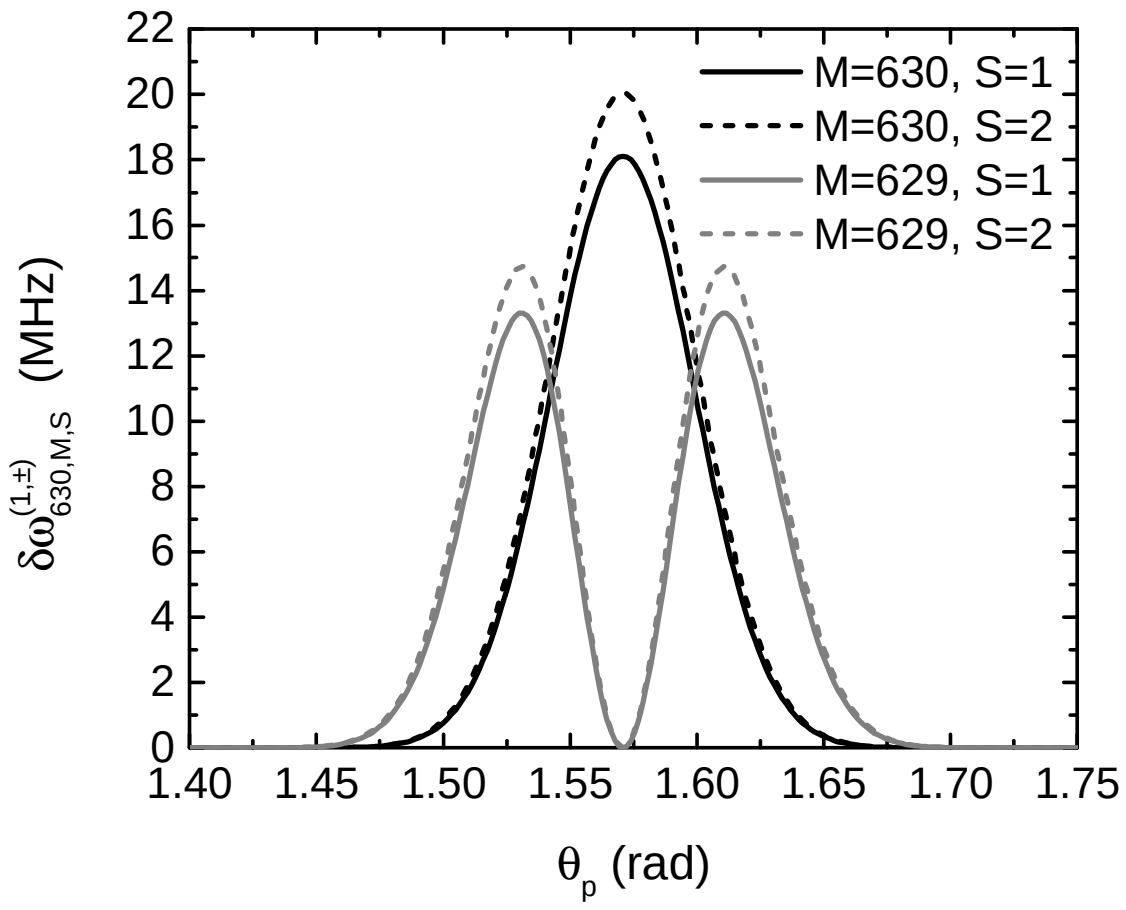

FIG. 4. Dependence of $\delta\omega_{L,L,S}^{(1,+)}$ and $\delta\omega_{L,L-1,S}^{(1,-)}$ on the particle's angular coordinate for $L$=630, $M$=$L$,$L$-1 and $S$=1,2

mode occurs at $\theta_p = \pi / 2$ (equatorial position), and it decreases very fast with deviation of the particle from the equatorial plane: the change of $\theta_p$ by only $5\%$ results in the order of magnitude decrease of the splitting. The $M = L-1$ mode shows the opposite behavior: the splitting vanishes at the equatorial position and reaches the maximum value at $\theta_p^{(\max)} = \tan^{-1}\sqrt{L-2}$ . Two other frequency shifts $\delta\omega_{L,L,S}^{(1,-)}$ and $\delta\omega_{L,L-1,S}^{(1,+)}$, while are not technically zeroes, are very small and are not shown. For such large values of $L$ they can be safely neglected so that the results in this case can again be discussed in terms of the frequency splitting used in most literature on this topic.

The graphs presented in Fig. 3 and 4 allow for comparison of our theoretical results with data presented in Ref. [19]. This comparison, however, is complicated by the lack of information about the angular coordinate of the particle $\theta_p$ interacting with the WGM. This difficulty is evident from the fact that the results of two different measurements, presented even in the same paper with nominally identical resonators and particles, do not agree with each other. Figure 3a of Ref. [19] shows splitting $19.4\,MHz$ for $R_r = 43.5\,\mu m$ and $R_p = 50\,nm$, while Figure 3b of the same paper for the same sizes of the resonator and particle shows splitting less than $10\,MHz$ . The dissimilarity most likely is due to different placements of the particle in these two experiments.

In order to use Eq. (18) to compute the amount of the frequency splitting for the TE mode for parameters given in Ref. [19] one has to make assumptions about position of the particle, the polar and radial orders of the excited WGM. We make these computations assuming the equatorial position of the particle and the fundamental $(M = L)$ polar nature of the mode. Then, if the radial order is chosen to be $S = 1$, we find that the splitting is equal to $18\,MHz$ . Presuming that the excited mode is of the second radial order we find the splitting to be equal to $20\,MHz$ . The experimental result presented in Figure 3a of Ref. [19] lays between these two values. This fact by itself validates the assumption that it is the $M = L$ mode, which is being excited in that experiment, because, as one can see from Fig. 4, the splitting in the $M = L-1$ mode is much smaller than the observed values for any position of the particle. Further, since the observed value of the splitting is larger than the one predicted for the 1st radial mode, we can conclude that the WGM really excited in this particular experiment was of the second radial order, and that the particle was positioned slightly off the equatorial plane. In principle, comparing the experimental splitting with our Fig. 4 one could surmise the angular position of the particle. The data presented in Fig. 3b of Ref. [19] show much smaller splitting, and, therefore, do not allow for unambiguous interpretation.

An interesting approach to measuring the angular position of the particle in sensing experiments was suggested in recent Ref. [35]. It was proposed that by comparing the modifications of the WGM resonances with $M = L$ and $M = L-1$ observed in a single experiment one can deduce the angular coordinate of the particle. Indeed, if one neglects the dependence of the radiative width $\gamma^{(1)}_{L,M,S}$ on $M$ , and assumes that both excited modes are of the same radial order, it can be derived from Eq.(18)

$$\frac{\delta\omega^{(1,-)}_{L,L-1,S}}{\delta\omega^{(1,+)}_{L,L,S}} = \frac{\left(d^{(L)}_{1,L-1} - d^{(L)}_{-1,L-1}\right)^2}{\left(d^{(L)}_{1,L} + d^{(L)}_{-1,L}\right)^2} = \frac{2\left(L-1\right)^2}{L}\tan^2\left(\theta_p - \pi/2\right). \tag{20}$$

Eq. (20) is a generalization of Eq. (5) of Ref. [35] valid for arbitrary orbital modal number and particle's angular coordinate $\theta_p$. While this proposal was verified experimentally in Ref. [35], these experiments were conducted in the regime of overlapping resonances. In this regime, in order to determine the spectral position of the resulting single resonance, one needs to take into account the entire frequency dependence of coefficients $c^{(1)}_{m,L}$ defined in Eq. (16), and direct comparison of Eq. (18) with the experimental data of Ref. [35] is not possible. This situation deserves a separate study, which is under way and will be presented in a subsequent publication.

Most of sensing experiments with WGMs, such as in Ref. [9–18,35], use WGMs with large values of the orbital number $L \approx 300 \div 600$. For such orbital numbers one can consider asymptotic behavior of Eq. (18) for $L \gg 1$ and present it in a simplified form containing only immediately available material parameters. To this end, we first present Eq. (18) for a dielectric particle characterized by real refractive index $n_p$ in the form

$$\delta\omega^{(1,\pm)}_{L,M,S} = -\frac{\chi}{2\pi}\left(d^{(L)}_{1,M} \pm d^{(L)}_{-1,M}\right)^2 \gamma^{(1)}_{L,M,S}\left(\frac{n_1\omega^{(1)}_{L,M,S}}{c}\right)^3 \left(2L+1\right)\left[y_L(n_1 k_{L,M,S} d_{pr})\right]^2, \tag{21}$$

where we expanded the particle's Mie-Lorentz coefficient $\alpha_p$ in terms of small parameter $k_1 R_p \ll 1$, neglected the small real part of the spherical Hankel function replacing it by the spherical Bessel function of the second kind, $y_L(z)$, used Eq. (10) for the translation coefficient $B^{(+)}_{1,L;1,1}(k_1, d_{pr})$, and introduced particle's polarization

$$\chi = 4\pi R_p^3 n_1^2 \frac{n_p^2 - n_1^2}{n_p^2 + 2n_1^2} . \tag{22}$$

Next, we replace the radiative linewidth of WGMs, $\gamma^{(1)}_{L,M,S}$, with its asymptotic expression derived in Ref. [43] for spherical resonators

$$\gamma^{(1)}_{L,M,S} = \frac{c^3}{\left(n_r^2 - n_1^2\right) n_1 \left[\omega^{(1)}_{L,M,S}\right]^2 R_r^3 \left[y_L(n_1 k_{L,M,S} R_r)\right]^2} . \tag{23}$$

In the case of the fundamental mode, Eq. (21) can now be presented as

$$\begin{gathered}
\frac{\delta\omega^{(1,+)}_{L,L,1}\left(\theta_p\right)}{\omega^{(1)}_{L,L,1}} = -\frac{L}{\left(L+1\right)\sin^2\theta_p}\left\{2\chi\left|Y_L^L\left(\theta_p,\varphi\right)\right|^2 \frac{1}{\left(n_r^2 - n_1^2\right)R_r^3}\left[\frac{y_L(n_1 k_{L,L,1} d_{pr})}{y_L(n_1 k_{L,L,1} R_r)}\right]^2\right\} \\
\frac{\delta\omega^{(1,-)}_{L,L,1}\left(\theta_p\right)}{\omega^{(1)}_{L,L,1}} = -\frac{L\cos^2\theta_p}{\left(L+1\right)\sin^2\theta_p}\left\{2\chi\left|Y_L^L\left(\theta_p,\varphi\right)\right|^2 \frac{1}{\left(n_r^2 - n_1^2\right)R_r^3}\left[\frac{y_L(n_1 k_{L,L,1} d_{pr})}{y_L(n_1 k_{L,L,1} R_r)}\right]^2\right\}
\end{gathered} , \tag{24}$$

where $Y_L^L\left(\theta_p,\varphi\right)$ is the standard scalar spherical harmonics. For the equatorial position of the particle $\delta\omega^{(1,-)}_{L,L,1}\left(\theta_p\right)$ vanishes, as expected. In the case of modes with large orbital numbers all frequency shifts are only appreciable for very small deviations of the particle's position from the equatorial plane: $\left|\theta_p - \pi/2\right| \le 1/\sqrt{L}$. For such angles the factor $L/\left[\left(L+1\right)\sin^2\theta_p\right]$ in the expression for $\delta\omega^{(1,+)}_{L,L,1}\left(\theta_p\right)$ is almost equal to one. It can be shown that the remaining expression inside the curly brackets $\{\cdots\}$ reproduces the result of the phenomenological theory of Ref. [4] when it is applied to a spherical resonator. The second shift $\delta\omega^{(1,-)}_{L,L,1}\left(\theta_p\right)$, which vanishes in the phenomenological theory [4] for any particle's position, while is not exact zero in our consideration, remains very small due to an extra factor $\left(\theta_p - \pi/2\right)^2 \le 1/\mathrm{L}$. Similar agreement between phenomenological results and our calculations is found for the modes with other polar numbers as well. For instance, for the mode with $M = L-1$, the main frequency shift $\delta\omega^{(1,-)}_{L,L-1,1}\left(\theta_p\right)$ obtained from Eq. (21) coincides with the frequency splitting of the phenomenological approach with accuracy up to the same factor of the order of unity, while the minor frequency shift given by $\delta\omega^{(1,+)}_{L,L-1,1}\left(\theta_p\right)$ can be shown to be small as $\delta\omega^{(1,+)}_{L,L,1}\left(\pi/2\right)/L$.

Thus, we can see that for WGM with large values of the orbital number $L$ the predictions of the phenomenological theory of Ref. [4] agree with the results of our *ab initio* calculations within experimentally available accuracy. This situation changes, however, in the case of smaller resonators supporting WGMs with orbital numbers $L \sim 30 \div 70$. We consider as an example resonators with radius $R_r = 2.5\,\mu m$ in air $\left(n_1 = 1\right)$ studied experimentally in Ref. [44]. Figure 5 shows angular dependence of both frequency shifts $\delta\omega^{(1,+)}_{L,L,1}\left(\theta_p\right)$ and $\delta\omega^{(1,-)}_{L,L,1}\left(\theta_p\right)$ for the fundamental $L = 40$ mode of the first radial order. The main feature of the presented graphs that should be noted is that the frequency shifts generated by the particle in such a small resonator are an order of magnitude larger than in the case of larger resonators presented in Fig. 3 and Fig. 4 despite the fact that the Q-factors in the former case is smaller. The main reason for this enhancement is that the particle on the surface of a smaller resonator feels a much stronger field due to its smaller distance from the center of the resonator. Such a strong enhancement, however, is only possible for a resonator in air, because placing it in an aqueous medium will degrade the Q-factors of the respective modes rendering them unusable. Still, one can raise a question about finding the smallest resonator usable for sensing in aqueous environments and about determining the optimal experimental conditions resulting in the maximum detection sensitivity. Such conditions would include the working wavelength, size and the refractive index of the resonator. Preliminary calculations showed that one can increase the amount of the shift produced by a particle in water by the factor of two or three by reducing the size of the resonator to $10-20\ \mu m$, but more detailed study of this issue lies outside of the scope of this paper.

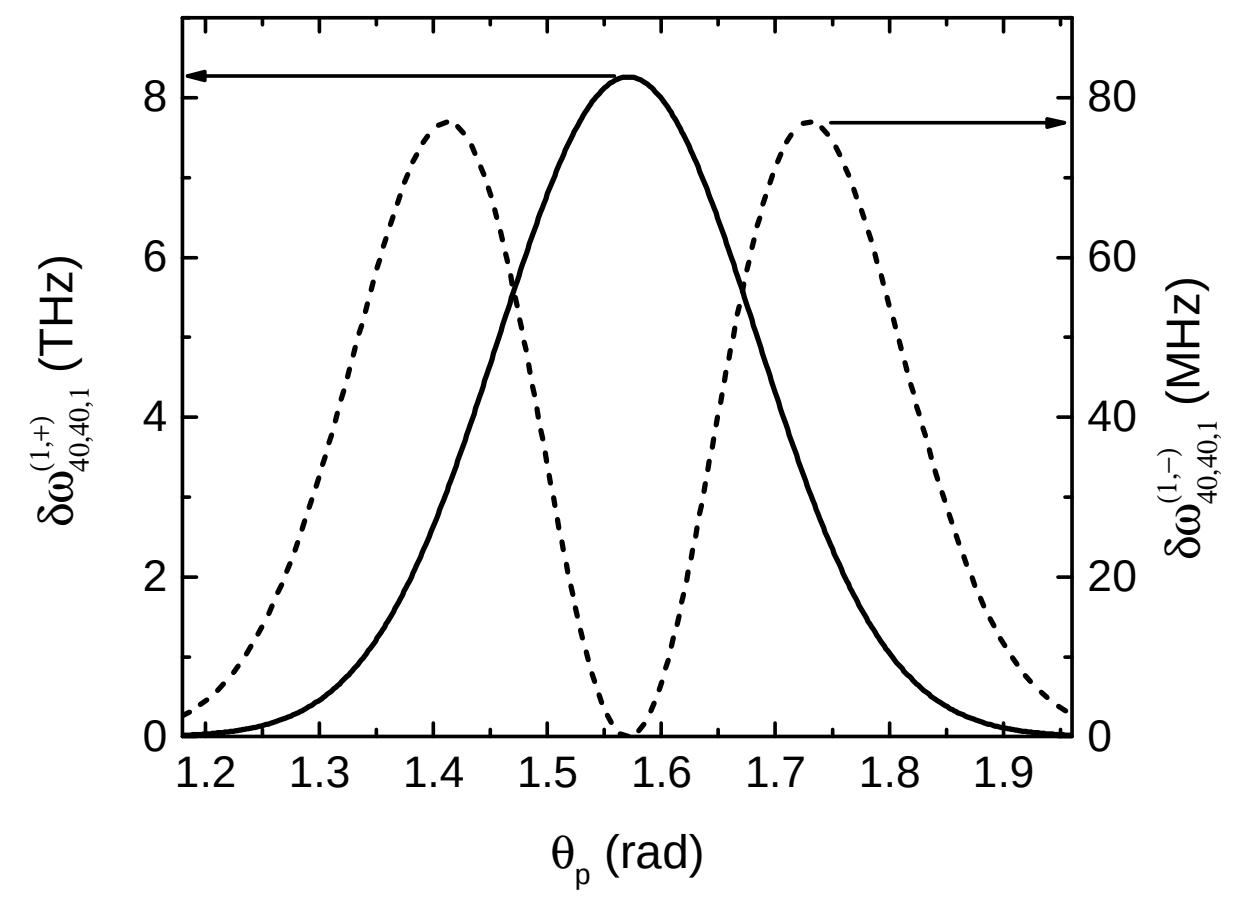

FIG. 5. Dependence of the major, $\delta\omega^{(1,+)}_{40,40,1}$, (left vertical axis) and minor, $\delta\omega^{(1,-)}_{40,40,1}$, (right vertical axis) frequency shifts on the angular coordinate of the particle for a fundamental $L = 40$ mode of the first radial order.

The second important feature of the graphs in Figure 5 is that while $\delta\omega_{L,L,1}^{(1,-)}\left(\theta_p\right)$ remains much smaller than $\delta\omega_{L,L,1}^{(1,+)}\left(\theta_p\right)$, it is still larger than the splitting observed for instance in Ref. [19], and is, therefore, observable. Observation of this shift would provide an additional reference points allowing to determine both position and the size of the particle from the observation of a single fundamental mode. Indeed, it follows from Eq. (18) that

$$\frac{\delta\omega_{L,L,1}^{(1,-)}}{\delta\omega_{L,L,1}^{(1,+)}}=\frac{\left[d_{1,L}^{(L)}\left(\theta_p\right)-d_{-1,L}^{(L)}\left(\theta_p\right)\right]^2}{\left[d_{1,L}^{(L)}\left(\theta_p\right)+d_{-1,L}^{(L)}\left(\theta_p\right)\right]^2} \tag{25}$$

without any additional approximations concerning the properties of the radiative decay rates $\gamma_{L,M,S}^{(1)}$. This approach should be contrasted with the method proposed in Ref. [35], which requires observation of two different WGMs and whose validity depends on the assumption that $\gamma_{L,M,S}^{(1)}\approx\gamma_{L,M-1,S}^{(1)}$ and that the two excited modes are of the same radial order.

B. TM polarization

Coefficients $p_\mu^{(2)}$ in Eq. (8) for the field scattered by the particle in the case of excitation of a TM polarized WGM are determined by matrix $U_{m,\mu}$, which depends on the translation coefficients $A_{m,\nu;m,1}^{(+)}(k_1,\mathbf{d}_{pr})$. Unlike $B_{m,\nu;m,1}^{(+)}(k_1,\mathbf{d}_{pr})$, these translation coefficients do not vanish at $m=0$ and, as a result, all three coefficients $p_0,\ p_{-1},\ p_1$ contribute to the field scattered by the resonator. One of the immediate consequences of this fact is that both resonances in the particle-induced doublet are red-shifted from the initial WGM even at the equatorial position of the particle. Frequencies and widths of these resonances are determined by the poles of the coefficients $g_{m,L}^{(2)}$ in the multipole expansion of the TM field scattered by the resonator, Eq.(1), which are found to be

$$g_{m\pm,L}^{(2)}=\frac{b_0\left(-i\right)^M\left(d_{m,M}^{(L)}\pm d_{-m,M}^{(L)}\right)}{\left(\tilde{T}_{M,L;M,L}^{(2)}\right)^{-1}-\alpha_p\left(\left(d_{1,M}^{(L)}-d_{-1,M}^{(L)}\right)^2\left[A_{1,L;1,1}^{(+)}\left(k_1,d_{pr}\right)\right]^2+2\left[d_{0,M}^{(L)}\right]^2\left[A_{0,L;0,1}^{(+)}\left(k_1,d_{pr}\right)\right]^2\right)}, \tag{26}$$

$$g^{(2)}_{m\mp,L} = \frac{(-i)^M b_0 \left(d^{(L)}_{m,M} \mp d^{(L)}_{-m,M}\right)}{\left(\tilde{T}^{(2)}_{M,L;M,L}\right)^{-1} - \alpha_p \left(d^{(L)}_{1,M} + d^{(L)}_{-1,M}\right)^2 \left[A^{(+)}_{1,L;1,1}\left(k_1, d_{pr}\right)\right]^2}, \tag{27}$$

where $g^{(2)}_{m\pm,L} = g^{(2)}_{m,L} \pm g^{(2)}_{-m,L}$. Subscript and respective signs "+" ("-") should be used with even (odd) values of polar number $m$ in Eq. (26) and odd (even) values of $m$ in Eq. (27). We can see that, similar to the case of TE polarization, all coefficients resonate at one of two different frequencies, which we again present in the form of their deviations $\delta\omega^{(2,-)}_{M,L,S}$ and $\delta\omega^{(2,+)}_{M,L,S}$ from the initial WGM resonance:

$$\begin{aligned}\delta\omega^{(2,-)}_{L,M,S} &= \left(d^{(L)}_{1,M} - d^{(L)}_{-1,M}\right)^2 \gamma^{(2)}_{L,M,S} \operatorname{Im}\left(\alpha_p \left[A^{(+)}_{1,L;1,1}(n_1 k_{L,M,S}, d_{pr})\right]^2\right) + \\ &\quad 2\left[d^{(L)}_{0,M}\right]^2 \gamma^{(2)}_{L,M,S} \operatorname{Im}\left(\alpha_p \left[A^{(+)}_{0,L;0,1}(n_1 k_{L,M,S}, d_{pr})\right]^2\right),\end{aligned} \tag{28}$$

$$\delta\omega^{(2,+)}_{L,M,S} = \left(d^{(L)}_{1,M} + d^{(L)}_{-1,M}\right)^2 \gamma^{(2)}_{L,M,S} \operatorname{Im}\left(\alpha_p \left[A^{(+)}_{1,L;1,1}(n_1 k_{L,M,S}, d_{pr})\right]^2\right). \tag{29}$$

Expressions for respective linewidths can be obtained from Eq. (28) and (29) by taking real part of the respective expressions instead of the imaginary part.

At the equatorial position of the particle $d^{(L)}_{1,M}(\pi/2) = (-1)^{L+M} d^{(L)}_{-1,M}(\pi/2)$ and one of the coefficients $g^{(2)}_{m\pm,L}$ vanishes. As a result, similar to the case of TE polarization, coefficients $g^{(2)}_{m,L}$ acquire definite parity $g^{(2)}_{m,L} = \pm g^{(2)}_{-m,L}$, which is different for even and odd polar numbers. Unlike the TE case, however, none of the frequency shifts given by Eq. (28) and (29) vanish at $\theta_p = \pi/2$ for fundamental $(M = L)$ mode. In this case (and all other cases, when $L - M$ is even), the first term in Eq. (28) vanishes resulting in equatorial frequency shifts given by

$$\begin{aligned}\delta\omega^{(2,-)}_{L,M,S}(\pi/2) &= 2\left[d^{(L)}_{0,M}(\pi/2)\right]^2 \gamma^{(2)}_{L,M,S} \operatorname{Im}\left(\alpha_p \left[A^{(+)}_{0,L;0,1}(n_1 k_{L,M,S}, d_{pr})\right]^2\right) \\ \delta\omega^{(2,+)}_{L,M,S}(\pi/2) &= 4\left[d^{(L)}_{1,M}(\pi/2)\right]^2 \gamma^{(2)}_{L,M,S} \operatorname{Im}\left(\alpha_p \left[A^{(+)}_{1,L;1,1}(n_1 k_{L,M,S}, d_{pr})\right]^2\right)\end{aligned}. \tag{30}$$

Comparing the translation coefficients $A^{(+)}_{1,L;1,1}(n_1 k_{L,M,S}, d_{pr})$, $A^{(+)}_{0,L;0,1}(n_1 k_{L,M,S}, d_{pr})$ and $B^{(+)}_{1,L;1,1}(n_1 k_{L,M,S}, d_{pr})$ one can see (Eq. (10)) that the frequency shifts of the TM modes are related to the shifts of the TE modes as $\delta\omega_{L,M,S}^{(2,+)}(\pi/2) < \delta\omega_{L,M,S}^{(1,+)}(\pi/2) < \delta\omega_{L,M,S}^{(2,-)}(\pi/2)$. The situation is different for modes characterized by odd values of $L-M$. In this case $d_{1,M}^{(L)}(\pi/2) + d_{-1,M}^{(L)}(\pi/2) = 0$ and the frequency shift $\delta\omega_{L,M,S}^{(2,+)}$ vanishes at $\theta_p = \pi/2$, and since $d_{0,M}^{(L)}(\pi/2) = 0$ for odd $L-M$, $\delta\omega_{L,M,S}^{(2,-)}$ acquires at $\theta_p = \pi/2$ a minimum value given by the same expression as the second line in Eq. (30).

The separation of the coefficients $g_{m,L}^{(2)}$ according to the parities of the polar number for the equatorial position of the particle can again be traced to the reflection symmetry of the system. This symmetry also helps to understand the difference between TE and TM modes. Indeed, unlike the TE situation, the field of TM polarization at the location of the particle has a component along the polar axis of the particle's system, which does not change sign upon reflection in the equatorial plane. It, therefore, remains different from zero if the field is described by even in $\varphi$ functions and vanishes for the odd in $\varphi$ field. In the latter case, however, the same arguments as in the TE case demonstrate that the component of the field perpendicular to the polar axis remains non-zero. Thus, the TM electric field has non-zero value at the location of the nanoparticle regardless of the parity of the polar numbers of the multipole contributions to it. As a result, the frequencies of resonances originating from coefficients $g_{m,L}^{(2)}$ with either odd or even $m$ are shifted from the single resonator value. The magnitude of the shift is, however, different for odd and even $m$, reflecting the fact that different Cartesian components of the field participate in the interaction with the particle in each of these cases.

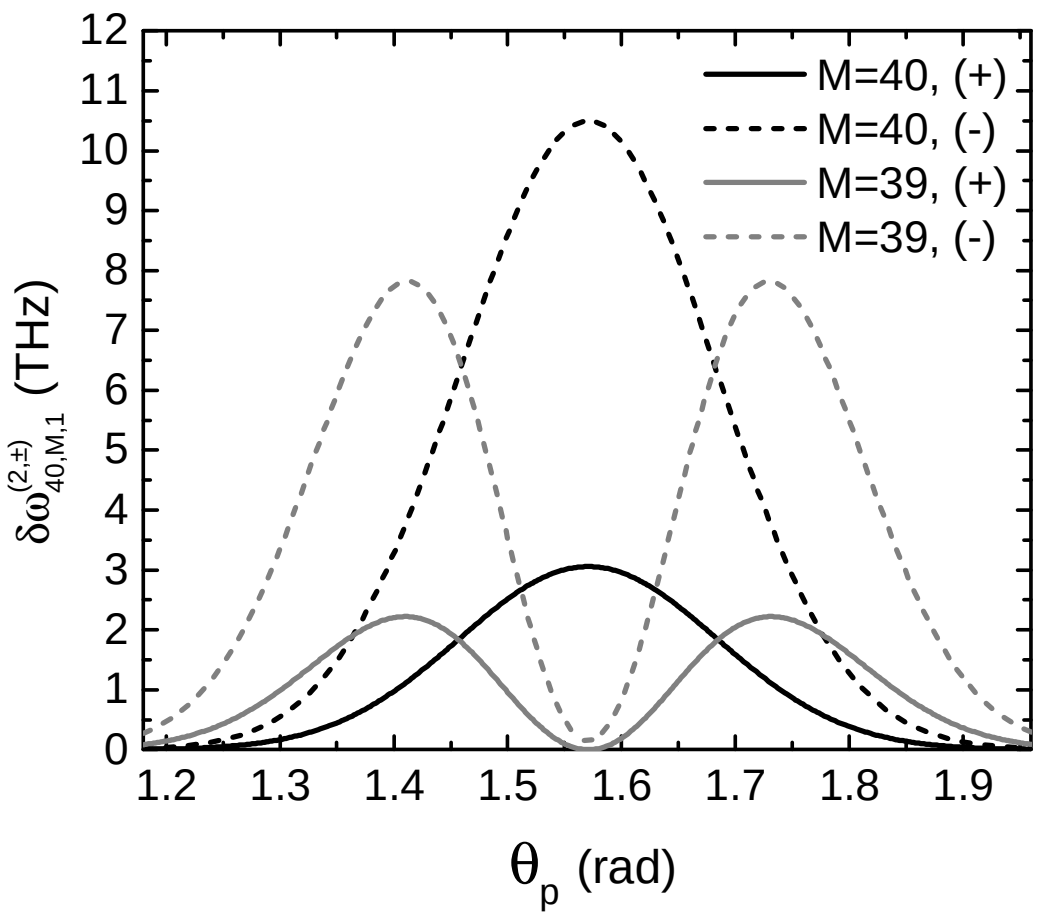

FIG. 6. Angular dependence of the two particle-induced frequency shifts $\delta\omega_{40,M,1}^{(2,-)}$ (dashed lines) and $\delta\omega_{40,M,1}^{(2,+)}$ (solid lines) for the $L$ = 40 TM modes of the first radial order. The black single-maximum lines present this dependence for the fundamental mode. The gray lines with two maxima present $M$ = 39 polar mode.

We illustrate the properties of the particle-induced frequency shifts for WGMs of TM polarization by plotting $\delta\omega^{(2,-)}_{L,M,S}$ and $\delta\omega^{(2,+)}_{L,M,S}$ as functions of the particle angular coordinate for two values of the orbital index $L$. Figure 6 presents the results of computations for $L = 40$ and two values of polar number $M = L$ (black lines) and $M = L-1$ (gray lines). The resonator and the particle are assumed to be in the air $(n_1 = 1)$. As expected, $\delta\omega^{(2,+)}_{40,40,1} < \delta\omega^{(1,+)}_{40,40,1} < \delta\omega^{(2,-)}_{40,40,1}$, while behavior of $\delta\omega^{(2,+)}_{40,39,1}$ is, to a large extent, similar to the behavior of the minor shift $\delta\omega^{(1,-)}_{40,40,1}$ of the TE fundamental mode. It is interesting that in this case even the minimum of $\delta\omega^{(2,+)}_{40,39,1}$ at $\theta_p = \pi/2$ remains experimentally observable.

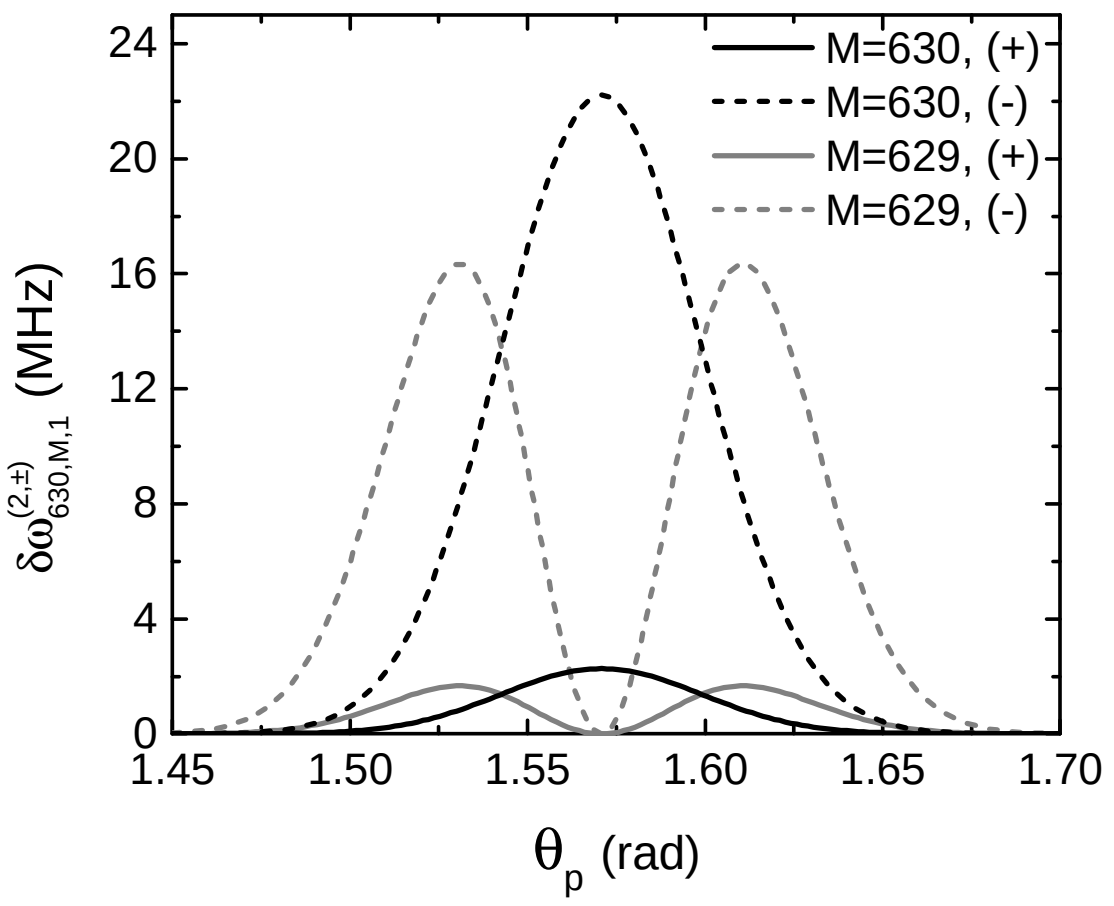

FIG. 7. The particle-induced frequency shifts $\delta\omega^{(2,-)}_{630,M,1}$ (dashed lines) and $\delta\omega^{(2,+)}_{630,M,1}$ (solid lines) versus particle's angular coordinate for the TM $L$=630 fundamental (black) and $M$=$L$-1 (gray) modes of the first radial order.

The situation does not change qualitatively for larger values of $L$, but quantitative difference is again rather striking. Figure 7 presents the same plots for $L = 630$ and $n_1 = 1.33$ (aqueous environment). The magnitude of the particle-induced frequency modifications have significantly decreased, but, unlike the case of TE polarization, both frequency shifts $\delta\omega^{(2,-)}_{630,M,1}$, $\delta\omega^{(2,+)}_{630,M,1}$ for $M = L$ and $M = L-1$ remain experimentally observable for a particle positioned in the proximity of the equatorial plane.

## IV. CONCLUSION

In this work we present results of the theoretical study of spectral modifications of WGMs of a spheroidal resonator induced by a small (subwavelength) particle adsorbed on the resonator's surface. We use T-matrix approach [33] to characterize the field scattered by the resonator and the dipole approximation for the field of the particle. We show that this system can be treated analytically in the so-called "resonant approximation" if one neglects elements of the T-matrix, which are non-diagonal in the polarization and orbital momentum indexes (these elements are responsible for coupling between WGMs of different polarizations and different angular momentums due to deviation of the resonator's shape from spherical). In addition, we assume that the particle couples only degenerate clockwise and counterclockwise WGMs, characterized by polar numbers differing in their sign only. This approximation describes well spheroidal resonators with moderate ellipticity: it is small enough to allow neglecting the non-diagonal elements of the T-matrix, but large enough so that the frequency splitting between WGMs with the same orbital number and different polar numbers is much larger than the particle-induced spectral modifications.

This approximation allowed us to derive relatively simple analytical formulas describing particle-induced modifications of WGM resonances of both TE and TM polarizations (since we neglect the cross-polarization scattering we can assign a definite polarization to the modes of the spheroidal resonator). Using these formulas we carried out a comprehensive analysis of the particle-induced effects in the spectrum of WGMs, which included comparison of the theoretical results with available experimental data and predictions of the earlier phenomenological treatments of this problem.

First of all, we found that even moderate deviation of the resonator's shape from spherical results in a significant decrease of the particle-induced spectral effects from values expected for ideally spherical resonators. This decrease is caused by lifting of the $2L+1$-fold degeneracy of WGM in spherical resonators, and as long as the spectral distance between resonances with the same $L$, but different polar numbers $M$, exceeds the particle-induced frequency shifts, the latter do not depend explicitly on the ellipticity of the resonator. This result explains a discrepancy between predictions of the *ab initio* calculations of Ref. [5,6] and experimental data. It also indicates that by reducing the ellipticity of the resonator below certain critical values one can hope to significantly increase the magnitude of the particle-induced frequency shifts. This issue will be discussed in more details in a separate publication.

Our calculations demonstrated that the spectrum of the resonator-particle system of both TE and TM polarizations is characterized by doublets of closely positioned resonances, both of which are red-shifted from the frequency of the respective initial single-resonator WGM. Both TM resonances originating from the fundamental mode (or any other mode with polar number $M$ such that $L-M$ is even) demonstrate a maximum particle-induced shift when the particle is in the equatorial plane of the resonator. The behavior of TE resonances is different: only one member of the TE doublet behaves similarly to its TM counterpart, while the second TE resonance moves toward the frequency of the initial WGM when the particle approaches the equatorial plane and merges with it at $\theta_p = \pi / 2$. These finding demonstrate that the preservation of the resonance at the single resonator frequency, which is one of the main results of Ref. [4], is not a universal feature of the WGM resonators. We traced it to the reflection symmetry of the resonator-particle system with respect to the equatorial plane of the resonator, which is destroyed, e.g. when the particle is shifted from the equatorial position. The results obtained in this paper also demonstrated that the third resonance for TM polarized modes predicted for spherical resonators in Ref. [5,6] is an artifact of the spherical approximation and disappears in moderately spheroidal resonators.

The derived expressions for the modified resonance frequencies agree very well with available experimental data reported, for instance, in Ref. [19] for WGMs of TE polarization characterized by relatively large orbital numbers $(L \approx 630)$. We also showed that in the limit $L \gg 1$, which is appropriate for these experiments, the predictions of the phenomenological theory of Ref. [4] agree asymptotically with our results for the modes of TE polarization. However, we also found that one can increase the particle-induced frequency shifts by two orders of magnitude by transitioning to smaller resonators with diameters of the order of $4\,\mu m$ supporting WGMs with smaller orbital numbers of the order $L \approx 20 \div 50$. In this case one can resolve both members of the TE particle-induced doublet and measure their deviations from the position of the initial WGM resonance. Such measurements can be used to measure both position and the polarizability of the particle from observation of the spectral modifications of a single WGM. This situation should be contrasted with the approach proposed in Ref. [35], where determination of the particle position depended on observation of frequency shifts in two $M = L$ and $M = L - 1$ modes. It should be noted, of course, that low $L$ resonances possess sufficiently large Q-factors only if there exists an adequate contrast of refractive indexes between the resonator and its surrounding. For typically used silica resonators it means that the measurements must take place in air. In aqueous environment, usually used for biosensing applications, WGMs with $L \approx 20 \div 50$ in such resonators become too broad to be useful. Nevertheless, one can expect to gain significant increase in the frequency shifts even in the aqueous environment by choosing the parameters of the resonator and experimental conditions allowing for use of smaller resonators supporting WGMs with orbital numbers of the order of $L \le 100$.

## ACKNOWLEDGEMENTS

Authors would like to express their deepest gratitude to Professor Steve Arnold for numerous illuminating discussions without which this paper would not have appeared.

This work was supported by the PSC-CUNY grant #66678-00 44.